\documentclass{article}

\usepackage{hyperref}
\usepackage{color}
\usepackage{graphicx}
\usepackage{amstext,amsmath,amssymb,amsfonts,bbm, amsthm}
\usepackage{fancyhdr}

\begin{document}

\title{On the AdS/CFT and its proof}

\author{Leonardo Ort\'{i}z\footnote{lortiz@fis.cinvestav.mx} \vspace{1cm}\\
Departamento de F\'{i}sica \\
Centro de Investigaci\'{o}n y de Estudios Avanzados del I. P. N. \\ 
Apdo. 14-740, CDMX, M\'{e}xico}

\maketitle

\begin{abstract}
In this letter we examine some structural aspects of the AdS/CFT such as the way of obtaining the expectation value of product of operators of the CFT, and ideas that should be considered when a proof of AdS/CFT is under consideration.  
\end{abstract}

\section{Introduction}

The AdS/CFT correspondence, as proposed by Maldacena \cite{mal98} and elaborated on by Witten \cite{wit98} and Gubser \textit{et}. \textit{al}. \cite{gub98}, has been without doubt a breakthrough in theoretical physics in the past century. This is understandable since if correct it would gives clues and advance our understanding of very interesting aspects of the still elusive theory of quantum gravity. The literature on this subject is vast, however a good introduction to it is given by MAGOO \cite{magoo00}, D'Hoker and Freedman \cite{ho02} and, for a more up to date lectures, Ramallo \cite{ra13}. For a more textbook oriented reader we recommend Ammon and Erdmenger \cite{am15} and N\v{a}stase \cite{na15}.

By now the AdS/CFT correspondence is enough developed that writing something new about it is challenging. The purpose of this work is to elaborate on an idea which once seen is almost obvious however no one so far has exposed. The idea is the following: the AdS/CFT as formulated by Witten and Gubser \textit{et}. \textit{al}. has a way to calculate correlation functions in the CFT from quantities in the bulk. This is done by disturbing the CFT by a source inherited from the bulk. On the other hand in a Lorentzian version of the AdS/CFT Bertola \textit{et}. \textit{al}. \cite{be00} and later Hamilton \textit{et}. \textit{al}. \cite{ha06}, \cite{ha206}, \cite{ha07}, \cite{ha08} can calculate the correlation functions of the CFT by a well defined limit. One can ask why there are two ways of calculating the correlation functions in the Lorentzian version of the AdS/CFT. The purpose of this letter is to point out that once one makes the correspondence proposed by Hamilton \textit{et}. \textit{al}. between the field on the boundary obtained as a limit of the field in the bulk and the operator on the boundary then both ways of calculating the correlation functions coincide eliminating any ambiguity.  

\section{Witten and Gubser et. al. formulation of the AdS/CFT}

In this section we will outline the way in which correlation functions of the CFT are calculated from quantities in the bulk. The basic idea is to identify the partition functions in the bulk and in the boundary \cite{ma14}:
\begin{equation}
Z_{\texttt{Gravity}}[\phi_{0}(x)]=Z_{\texttt{Field Theory}}[\phi_{0}(x)]=\langle e^{\int{d^{d}x\phi_{0}(x)\textsl{O}(x)}}\rangle.
\end{equation}
Then we can calculate the correlation functions on the CFT by
\begin{equation}
\langle\textsl{O}(x_{1})...\textsl{O}(x_{n})\rangle=\frac{\delta}{\delta\phi_{0}(x_{1})}...\frac{\delta}{\delta\phi_{0}(x_{n})}Z_{\texttt{Gravity}}[\phi_{0}(x)].
\end{equation}
It is important to notice that in this approach we are deforming the CFT by the source $\phi_{0}(x)$. 

\section{Other proposals of the AdS/CFT}

It is fair to say that most of the research done in the AdS/CFT followed from Witten and Gubser \textit{et}. \textit{al}. and is done by people in string theory, however there are other approaches which deserve some mention and which after all they are not too different.

First, we would like to highlight the work of Bertola \textit{et}. \textit{al}. \cite{be00}. In this work it was shown how, at least for the scalar real field, to obtain a conformal theory in the boundary of AdS from a Wightman theory in the bulk. This works can be seen as one way AdS/CFT, AdS$\rightarrow$CFT.

Second, we mention the work of Hamilton \textit{et}. \textit{al}. \cite{ha06}, \cite{ha206}, \cite{ha07}, \cite{ha08}. In these works it was shown how to recover bulk expectation values from data in the boundary. In particular it was done in a great generality in two dimensions. Clearly this is a one way AdS/CFT, AdS$\leftarrow$CFT.

Clearly, these two approaches to the AdS/CFT are complementary. Hence we have a true (both directions) AdS/CFT. However we recall we are working here with the undeformed CFT.

\section{After all similar languages}

In the Lorentzian formulation of the AdS/CFT it is necessary to keep both the unnormalized and normalized modes in the bulk \cite{ba99}. So if we use both ways, Witten \textit{et}. \textit{al}. and Bertola \textit{et}. \textit{al}., to calculate expectation values in the boundary it seems we have two ways of obtaining the same object. However if now we make the identification suggested by Hamilton \textit{et}. \textit{al}. 
\begin{equation}
\phi_{0}\leftrightarrow\textsl{O},
\end{equation}
where $\phi_{0}$ is the bulk field restricted to the boundary and \textsl{O} is the field of the CFT. Then both ways of calculating the correlation functions in the CFT coincide and there is no contradiction.

We are glad this happens otherwise work on the AdS/CFT correspondence as for example \cite{ke99} would be in contradiction with old work on quantum fields on black holes such as \cite{li94}. See \cite{or13} or the following lines.

On general grounds, in the context of the AdS/CFT correspondence, one expects that a black hole sitting inside AdS should be dual to a thermal state on the conformal boundary of the black hole. This has indeed been obtained for the BTZ black hole by Keski-Vakkuri \cite{ke99}. In this work the Witten prescription \cite{wit98} to get the state on the boundary was used and the normalizable modes were not taken into account. However from the analysis of Balasubramanian \textit{et al.} \cite{ba99} one knows that the normalizable modes are indeed important in the AdS/CFT correspondence, since they are dual to states on the conformal boundary. In this section we show that the normalizable modes contribute with a similar amount to the state on the boundary under the prescription of the Boundary-limit Holography. This in particular shows a relation of past work on QFT in BTZ black holes with the AdS/CFT correspondence. As far as we know no one else before has pointed out this relation. The fact that the normalizable modes contribute on equal footing to the state on the boundary as the non-normalizable modes has been implicitly done in \cite{lOrt13}. However in \cite{li94} we did QFT in AdS then went to the boundary by using the Boundary-limit Holography \cite{be00} then took the restriction to the exterior of the BTZ black hole. In the present section we follow a different procedure, we have QFT in AdS then QFT in the BTZ black hole and the go to the boundary under the prescription of the Boundary-limit Holography \cite{be00} applied to the BTZ black hole. This suggests that, in certain cases, the following two procedures are equivalent:
\begin{equation}\nonumber
\textrm{QFT in bulk AdS}\rightarrow \textrm{Boundary-limit} \rightarrow \textrm{Restriction to the BTZ black hole}
\end{equation}
\begin{equation}\nonumber
\textrm{QFT in bulk AdS}\rightarrow \textrm{Restriction to the BTZ black hole}\rightarrow \textrm{Boundary-limit}
\end{equation}

In both procedures we get a thermal state given by
\begin{equation}\label{E:1}
F_{b}(\Delta
t,\Delta\phi)\sim\sum_{n\in\mathbb{Z}}\frac{1}{(\cosh\kappa\ell(\Delta\phi+2\pi
n)-\cosh\kappa(\Delta t-i\epsilon))^{\frac{1}{2}}},
\end{equation}
where $\kappa$ is the surface gravity, $\ell$ is related with the cosmological constant as $\Lambda=-\frac{1}{\ell^2}$ and $\Delta t=t_{1}-t_{2}$, $\Delta \phi=\phi_{1}-\phi_{2}$ are the coordinates inherited from the spacial BTZ black hole coordinates.

This is telling us that a state on AdS, the global vacuum, induces a state on its conformal boundary which in turn induces a state on the BTZ black hole; and this is equivalent to have the same state on AdS spacetime then a state on the BTZ black hole then a state on the conformal boundary of the BTZ black hole

Now we show that we get (\ref{E:1}) by two distinct procedures for the massless conformal real scalar field. This is done by taking into account our previous work \cite{lOrt13} and the work of Lifschytz-Ortiz \cite{li94}.

From (61) in \cite{lOrt13} we see that if $p=-\frac{1}{2}$ then we get (\ref{E:1}). In this case (61) in \cite{lOrt13} was calculated with the following procedure: QFT in bulk AdS, in Poincar\'{e} coordinates, then the Boundary-limit \cite{be00}, and then restriction to the exterior of the BTZ black hole in the boundary. We see that $p=-\frac{1}{2}$ corresponds to take the massless conformal scalar field. In this case we are assuming the global vacuum is the same that the Poincar\'{e} vacuum. This is a valid assumption since it has been shown \cite{uhDan99} that for most purposes they can be considered equivalent.

Now let us discuss the other procedure. According to \cite{li94} the two point function for the massless conformal real scalar field on the exterior of the BTZ black hole is given by
\begin{equation}\nonumber
G(x,x')=\sum_{n\in\mathbb{Z}}\frac{1}{\left(\frac{rr'}{r_{+}^2}\cosh\kappa\ell(\Delta\phi+2\pi
n)-1-\frac{(r^2-r_{+}^2)^\frac{1}{2}(r'^2-r_{+}^2)^{\frac{1}{2}}}{r_{+}^2}\cosh\kappa(\Delta t-i\epsilon)\right)^{\frac{1}{2}}}.
\end{equation}
Now we take the boundary limit of this expression. In order to do this we multiply by $\frac{(rr')^{\frac{1}{2}}}{r_{+}}$ and take the limit when $r,r'\rightarrow \infty$. In this limit we get (\ref{E:1}) again. In this case we have: QFT in bulk AdS spacetime, then restriction to the exterior of the BTZ black hole and then Boundary-limit. So we have two procedures to get the same state. These two procedures use only QFT in curved spacetime techniques and the Boundary-limit Holography ideas \cite{be00}.

\section{AdS$\rightarrow$CFT and AdS$\leftarrow$CFT in two dimensions}

In two dimensions it is easier to formulate the above mentioned ideas. See for example Hamilton \textit{et}. \textit{al}. works. Seeing together these ideas with Bertola \textit{et}. \textit{al}. work is clear that we have a real bijection bulk-boundary. It would be interesting to see if this bijection can be proven for other fields. Clearly this would be a direction to prove the AdS/CFT.   

\section{Conclusions}

We have discussed important structural aspects of the AdS/CFT correspondence. It is clear that the discussion in this letter is necessary otherwise we can fall in contradictions or ambiguities. Perhaps for many readers this letter is obvious however we feel it is necessary to spell out the details exposed here to avoid misunderstandings.\\

\vspace{0.5cm}
\textbf{Acknowledgments}: We thank Prof. Nora Bret\'{o}n for all her support and confidence. This work was sponsored by CONACYT-Mexico through a postdoctoral fellowship.\hspace{0.5cm}\\

After posting the previous version of this letter to the arXiv M. B. Fr\"{o}g pointed out that similar conclusions had been done before in \cite{mD02}. However when I was a Ph. D. student I did not understand the conclusions of this paper, which I knew at that time.

\end{document}